# Multi-resonant piezoelectric shunting induced by digital controllers for subwavelength elastic wave attenuation in smart metamaterial

**Gang Wang**[1], **Jianqing Cheng**[1], **Jingwei Chen**[2] **and Yunze He**[3]

[1] State Key Laboratory of Advanced Design and Manufacturing for Vehicle Body, Hunan University, 410082 Changsha, China

[2] College of Mechanical and Vehicle Engineering, Hunan University, 410082 Changsha, China

[3] College of Electrical and Information Engineering, Hunan University, 410082 Changsha, China

E-mail: zaehwang@hotmail.com, chenjingwei@hnu.edu.cn

## Abstract

Instead of analog electronic circuits and components, digital controllers that are capable of active multi-resonant piezoelectric shunting are applied to elastic metamaterials integrated with piezoelectric patches. Giving thanks to the introduced digital control technique, shunting strategies with transfer functions that can hardly be realized with analog circuits is possible now. As an example, the "pole-zero" method is developed to design single- or multi-resonant bandgaps by adjusting poles and zeros in the transfer function of piezoelectric shunting directly. Large simultaneous attenuations in up to three frequency bands at deep subwavelength scale (with the normalized frequency as low as 0.077) are achieved. The underlying physical mechanism is attributed to the negative group velocity of flexural wave within bandgaps. As digital controllers can be readily adapted via wireless broadcasting, the bandgaps can be tuned easily instead of tuning the electric components in analog shunting circuits one by one manually. The theoretical results are well verified experimentally with the measured vibration transmission properties where large insulations of up to 20dB in low-frequency ranges are observed.

## 1. Introduction

In the last decade, acoustic/elastic metamaterials (EMMs) have gained increasing interests due to their unnatural elastic properties. As a kind of structural material, EMMs can be designed to have significant ability of suppressing, absorbing or manipulating elastic waves by tailoring its metastructure at subwavelength scale. Thus a large structure is no longer necessary in order to gain a low-frequency bandgap,[1] which is helpful for low-frequency isolation or attenuation of sound and vibration,[2–7] negative refraction,[8] acoustic imaging,[9,10] and acoustic cloaking[11], etc.

Due to weak attenuations in locally resonant (LR) bandgaps[2] and the lack of real-time tuning, piezoelectric (PZT) shunting was introduced into the design of the EMMs.[12] The benefits are the lighter weight and the tunable abilities. Airoldi *et al*. designed tunable metamaterial beam with periodic arrays of resonant shunting (RS) of PZTs[13], and also proposed bi-RS to generate two LR bandgaps in it.[14] We also performed some theoretical and experimental investigation on EMM beams with passive PZT shunting[15,16]. This strategy has also been extended to flat plate hosting periodic arrays of shunted PZT patches. The elastic wave in the metamaterial thin plates[17–21] as well as the sound transmission loss through it[22] were both researched. Moreover, Casadei *et al*. implemented experimentally a tunable elastic wave waveguide by placing 8 PZT disks in the L-shaped waveguide of a two-dimensional phononic plate, where all the PZT disks were shunted by passive inductive circuits.[23] Kwon *et al*. also introduced the passive PZT shunting in the design of vibroacoustic metamaterials that can exhibit acoustic waveguides with tunable bandgaps.[24] Another research work by Nouh *et al.* demonstrated a metamaterial plate with tunable local resonators composed of piezo-membranes with adjustable stiffness controlled by external voltages.[25]

The passive RS technique can hardly produce large attenuation of vibration at deep subwavelength frequencies. As a revision of passive RS, negative capacitance PZT shunting has been introduced into the EMM in order to enhance the coupling effect.[26,27] It was also used in the design of cellular metamaterials that can exhibit reconfigurable, highly focused, subwavelength wave patterns theoretically.[28] However, this technique requires the tuning of the circuit very close to the stability limit in order to get the desired performance.[29] Recently, we proposed two different kinds of enhanced RS, i.e. the "resonator-amplifier"[30] and the "amplifier-resonator"[31] active strategies in order to intensify the resonant effect of the passive RS, and large attenuations within low-frequency ranges were observed.

All the above techniques are based on analog electronic components such as inductors, resistors and capacitors. Large inductors without internal resistance are required, which are usually realized by using synthetic circuits composed of 2 (Antoniou circuit that needs to be partly grounded) or 4 (floating inductor) operational amplifiers and some other components. For the multi-resonant shunting technique,[14] this situation becomes worse because $N \times N$ floating inductors are needed to gain a *N*-resonant shunting. In order to achieve two combined or separate band gaps, Zhou et al. proposed a high-order resonant shunt circuit that can exhibit two local resonances, and demonstrated the effects via simulations.[32] In their high-order resonant shunt circuit, the additional resonance was introduced by adding a capacitor, a resistor and a floating inductor to the traditional RL-shunt. The floating inductor will be the main complexity in practice too. Besides the complexity and the attendant inconvenient tunability, another shortage of the analog shunting circuits is the difficulty in design. Ordinarily, they are built by trial and error.

When linearity is assumed, all the dynamic behavior of the shunted PZT can be described as a SISO linear system. So it will be much more convenient if it's possible to bypass complex procedures related to analog circuits and design its transfer function directly. This paper is trying to make the first attempt in this region. For the first time, digital controllers (active shunting units) are designed and applied on an EMM structure composed of aluminum alloy beam and arrays of PZT patches glued on it. A "pole-zero" method is developed to design single- or multi-resonant transfer functions that related to the equivalent Young's modulus of the shunted piezoelectric patches. The transfer function is transformed into recurrence equations in order to be realized by digital controllers. Finally, both the single- and multi-resonant bandgap behaviour of the proposed smart EMM is analysed theoretically and validated experimentally. Compared with aforementioned works (including ours[30,31]) that based on analog shunting techniques, the main novelty and original contribution of this paper is as follows: Giving thanks to the introduced digital control technique, all the restrains that relate to analog electronic circuits can be cast off. New shunting strategies with transfer functions that can hardly be realized with analog circuits is possible now, which provides a new vision in the design of smart EMM. As an example, the proposed "pole-zero" method can be used to design the multi-resonant bandgaps directly by simply adjusting poles and zeros in the transfer function that relate to certain equivalent Young's module, instead of entwining into a confusing mass of analog circuits. Moreover, the proposed digital controllers can be readily adapted via wireless broadcasting, thus the bandgaps can be tuned easily instead of tuning electric components manually one by one. This advantage will be helpful in the design of self-adaptive smart metamaterials.

## 2. Theoretical model and experimental setup

*2.1. Theoretical model of the beam with digital shunting controllers*

Figure 1(a) illustrates the mechanical structure of a 6-period specimen of the proposed smart EMM beam, while figures 1(c) and 1(d) show the close-up views of it from different angle. The specimen is composed of an aluminum alloy beam and 6 pairs of PZT patches periodically glued on both sides of it. In each period, the beam's segment with the PZT patch is denoted by A, while the other is denoted by B. For the specimen with finite size, the segment on both ends of the beam are denoted by C. Each pair of PZT patches are placed with opposite polarizing directions along the $z$-axis and connected with a digital controller for active piezoelectric shunting, as shown in figure 1(b). All active shunting units are same and work independently under the same mechanism in order to guarantee the periodicity of the smart EMM structure. Each active shunting unit is composed of a charge amplifier, an analog-to-digital converter (ADC), a microprocessor, a digital-to-analog converter (DAC) and a proportional amplifier. The mechanical strain on PZT patch $a$ (the sensor patch) is converted into a voltage signal $V_1$ by the charge amplifier, and then digitalized by the ADC. Afterwards, the microprocessor uses these data to calculate in real-time the value of output signal $V_2$ based on the desired transfer function. Then the digital value is transferred into the analog voltage signal $V_2$ by the DAC. Finally, the voltage signal $V_2$ is magnified $\beta$ times with a proportional amplifier circuit and applied on the PZT patch $b$ in order to achieve the active PZT shunting based on the designed transfer function. The geometric and material parameters of the proposed smart EMM beam are listed in tables 1 and 2.

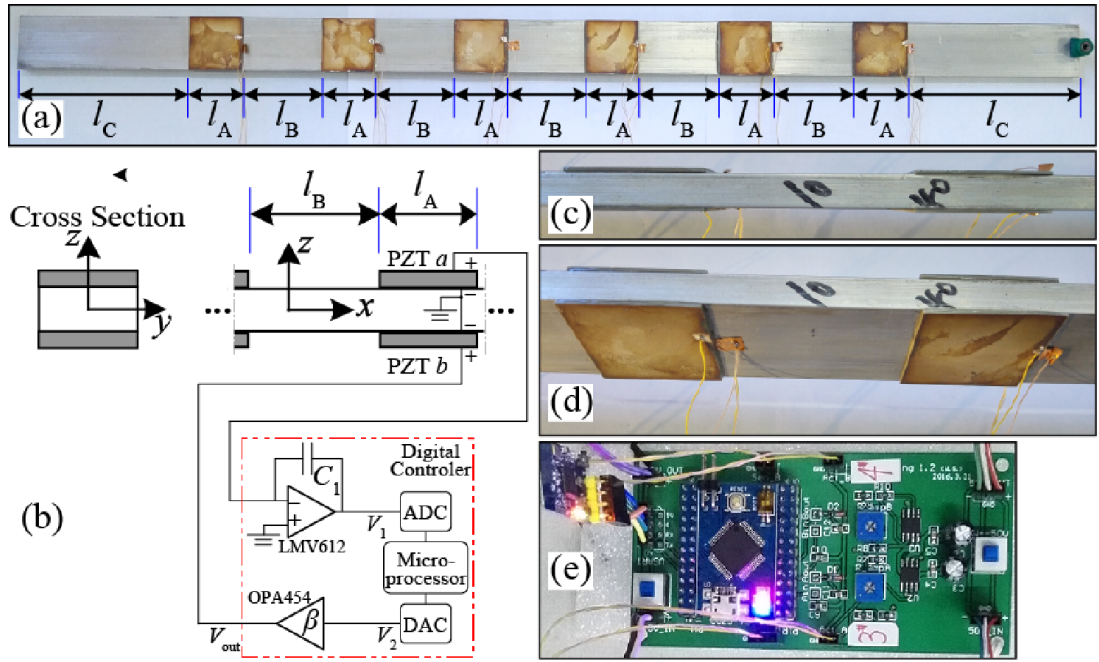

Figure 1. (a) Photo of the smart elastic metamaterial beam with 6 periods. (b) Sketch of one unit cell and the corresponding digital controller. (c) Close-up top view of the smart elastic metamaterial beam. (d) Close-up view of the specimen from oblique angle. (d) Photo of one circuit board that consists two independent digital controllers.

**Table 1.** Geometric and material parameters of the hosting beam (aluminum alloy 6061-T6)

| Parameter | Symbol | Value |
|---|---|---|
| Density | $\rho_{beam}$ | 2700 kg/m$^3$ |
| Young's modulus | $E_{beam}$ | 69.5 GPa |
| Width | $b_1$ | 0.04 m |
| Thickness | $h_1$ | 0.01 m |
| Length of segment A | $l_A$ | 0.04 m |
| Length of segment B | $l_B$ | 0.06 m |
| Mass of sensor | $m$ | 8 g |
| Number of periods | $N$ | 6 |
| Length of segment C | $l_C$ | 0.13 m |

**Table 2.** Geometric and material parameters of the piezoelectric patch (P-42)

| Parameter | Symbol | Value |
|---|---|---|
| Density | $\rho_{beam}$ | 7600 kg/m$^3$ |
| Compliance coefficient | $s_{11}^E$ | 11×10$^{-12}$ m$^2$/N |
| Piezoelectric strain constant | $d_{31}$ | -101×10$^{-12}$ C/N |
| Dielectric constant | $\varepsilon_{33}^T$ | 1.2390×10$^{-8}$ F/m |
| Width | $b_p$ | 0.04 m |
| Thickness | $h_p$ | 1×10$^{-3}$ m |

Figure 1(d) shows the photo of the low-cost circuit board that contains two independent digital controllers used for active PZT shunting in this paper. When transfer functions of different controllers are same, the multi-physics mechanism in each unit cell are the same and thus the periodicity is guaranteed. Two kinds of precise operational amplifiers (LMV612 and OPA454) from Texas Instruments are chosen to make each circuit board. One LMV612 is used as a two-channel charge amplifier that transfers the mechanical strain (in form of electric charge) of PZT patch *a* into a voltage signal; and two OPA454 capable of operating up to ±50V are chosen as two *β*-times amplifiers in order to provide high driving voltages applied on PZT patches *b*. The STM32F103RCT6 from STMicroelectronics that has two channels of DACs and several channels of ADCs on chip is used as the microprocessor and the ADC/DAC. ZigBee wireless units are connected with each circuit board and the computer in order to modify the parameters of the controllers synchronously.

As PZT patch *a* in figure 1(b) is equivalently grounded, the piezoelectric equations in it can be reduced[31]. Thus the relationship between the mechanical strain $S_1^a$ and voltage $V_1$

can be deduced as

$$V_1 = \frac{A_s d_{31}}{s_{11}^E C_1} S_1^a = F S_1^a \tag{1}$$

where $F = A_s d_{31} / s_{11}^E C_1$; $A_s$ is the area of the PZT electrode; $d_{31}$ is the piezoelectric strain constant; $s_{11}^E$ is the compliance coefficient of the PZT material at the constant electric field intensity; $C_1$=100nF is the capacitor in the charge amplifier circuit; and $S_1^a$ is the mechanical strain in PZT patch *a*.

As the voltage $V_1$ is in a constant ratio with the strain $S_1^a$, it can be used as a perfect sensor of the strain inside the PZT patch *a*. Furthermore, the effective elastic modulus $E_p^a$ of PZT patch *a* is a constant[31], which is the mechanical stress $T_1^a$ divided by the strain $S_1^a$ as

$$E_p^a = \frac{T_1^a}{S_1^a} = \frac{1}{s_{11}^E} \tag{2}$$

The voltages on both ends of the *β*-times amplifier have the relation as

$$V_{out} = \beta V_2 \tag{3}$$

where *β* is the amplifying ratio of the proportional amplifier circuit which is used to amplify the output voltage of the on-chip DAC to the output voltage of the OPA454 (the operational amplifiers). As the maximum output voltage ranges of the on-chip DAC and the OPA454 are ±1.65V and ±50V respectively, and OPA454's maximum voltage range has a little shrinking of up to 5% based on different current, the amplifying ratio *β* is chosen as *β*=100/3.3×95%≈28.8 in this paper.

Assuming the analog transfer function of the microprocessor along with ADC and DAC is $H_{DP}(s) = V_2/V_1$, we can have the following relationship by combining equation (1) with equation (3)

$$\frac{V_{out}}{S_1^a} = \beta F H_{DP}(s) \tag{4}$$

where *s* is the complex frequency variable in Laplace transform.

As the electric field in PZT patch *b* is $E_3^b = V_{out} / h_p^b$, alone with equation (4), the piezoelectric equations[31] in patch *b* can be transformed as

$$S_1^b = s_{11}^E T_1^b + \frac{d_{31} \beta F H_{DP}(s)}{h_p^b} S_1^a \tag{5}$$

where $h_p^b$ is the thickness of PZT patch *b*; $S_1^b$ and $T_1^b$ are the mechanical strain and stress of PZT patch *b* respectively.

When deep subwavelength flexural waves are considered, the strains of PZT patches *a* and *b* in one unit cell can be approximatively regarded as $S_1^b = -S_1^a$. Thus the equivalent elastic modulus of PZT patch *b* can be derived as mechanical stress $T_1^b$ divide by strain $S_1^b$

$$E_p^b(s) = \frac{T_1^b}{S_1^b} = \frac{h_p^b + d_{31}\beta F H_{\mathrm{DP}}(s)}{h_p^b s_{11}^E} \tag{6}$$

Assuming $s=j\omega$, PZT patches *b* can be considered as ordinary elastic materials with equivalent complex stiffness determined by equation (6) at certain frequencies.

If each segment A or B of the beam is considered as a Timoshenko beam, the flexural elastic waves in it are governed by the following differential equations

$$\frac{EI}{\rho S}\frac{\partial^4 u}{\partial x^4} - \frac{I}{S}\left(1+\frac{E}{\kappa G}\right)\frac{\partial^4 u}{\partial x^2 \partial t^2} + \frac{\partial^2 u}{\partial t^2} + \frac{\rho I}{\kappa G S}\frac{\partial^4 u}{\partial t^4} = 0 \tag{7}$$

where $u(x,t)$ is the transverse displacement of the beam along the *z*-axis, *E* and *G* are the Young's and shear modulus, respectively, $\rho$ is the density, *S* is the cross-section area, $\kappa$ is the Timoshenko shear coefficient, and *I* is the moment of inertia with respect to the axis perpendicular to the beam axis.

If harmonic vibrations are assumed as $u(x,t) = U(x)e^{i\omega t}$, where $\omega=2\pi f$, *f* being the frequency, the solution of Timoshenko equation (7) can be written as:[33]

$$U(x) = \begin{bmatrix} k_1^{-3}e^{k_1 x} & k_2^{-3}e^{k_2 x} & k_3^{-3}e^{k_3 x} & k_4^{-3}e^{k_4 x} \end{bmatrix}\boldsymbol{\psi} \tag{8}$$

where $k_j = (-1)^{[j/2]}\sqrt{[\alpha + (-1)^j \sqrt{\alpha^2+4\beta}]/2}$ (for j=1,...,4) depends on the mechanical and geometrical properties of the beam through $\alpha = -\rho\omega^2/E - \rho\omega^2/\kappa G$ and $\beta = \rho S\omega^2/EI - \rho^2\omega^4/E\kappa G$, [*j*/2] is the largest integer less than *j*/2, and $\boldsymbol{\psi} = \{A,B,C,D\}^\mathrm{T}$ are factors need to be determined.

In period *n*, the continuity of displacement, rotating angle, bending moment and shearing force at the boundary of beam segments A and B can be described as

$$\mathbf{H}_\mathrm{A}(l_\mathrm{A})\boldsymbol{\psi}_{\mathrm{A},n} = \mathbf{H}_\mathrm{B}(0)\boldsymbol{\psi}_{\mathrm{B},n} \tag{9}$$

While the continuity at the boundary of beam segment B in period *n* and beam segment A in period *n*+1 ensures

$$\mathbf{H}_\mathrm{B}(l_\mathrm{B})\boldsymbol{\psi}_{\mathrm{B},n} = \mathbf{H}_\mathrm{A}(0)\boldsymbol{\psi}_{\mathrm{A},n+1} \tag{10}$$

where

$$\mathbf{H}_{A,B}(l) = \begin{bmatrix} k_1^{-3}e^{k_1x} & k_2^{-3}e^{k_2x} & k_3^{-3}e^{k_3x} & k_4^{-3}e^{k_4x} \\ k_1^{-2}e^{k_1x} & k_2^{-2}e^{k_2x} & k_3^{-2}e^{k_3x} & k_4^{-2}e^{k_4x} \\ EIk_1^{-1}e^{k_1x} & EIk_2^{-1}e^{k_2x} & EIk_3^{-1}e^{k_3x} & EIk_4^{-1}e^{k_4x} \\ EIe^{k_1x} & EIe^{k_2x} & EIe^{k_3x} & EIe^{k_4x} \end{bmatrix}_{A,B} \quad (11)$$

Combine (9) and (10), we can get

$$\psi_{A,n+1} = \mathbf{T}\psi_{A,n} \quad (12)$$

where $\mathbf{T} = [\mathbf{H}_A(0)]^{-1}\mathbf{H}_B(l_B)[\mathbf{H}_B(0)]^{-1}\mathbf{H}_A(l_A)$ is the transfer matrix between adjacent periods of the beam.

For infinite period structure illustrated in figure1, Bloch boundary condition[34] on the beam can be described as

$$\psi_{A,n+1} = e^{iqa}\psi_{A,n} \quad (13)$$

where $qa$ is the wave propagation constant in the infinite periodic structure, $a=l_A+l_B$ is the lattice constant of the 1D metamaterial. Here $l_A$ and $l_B$ are the lengths of beam segments A and B.

Combining (13) with (12), a eigenvalue equation as below can be derived[31]

$$|\mathbf{T} - e^{iqa}\mathbf{I}| = 0 \quad (14)$$

For a given $\omega$, we can calculate the corresponding propagation constant $qa$ with (14). The real part of $q$ is the wave vector (also called the wave number in one-dimensional case). The graph of wave vector versus frequency is called the dispersion curve. When the propagation constant $qa$ is a real number, a flexural wave can propagate freely in the infinite metamaterial structure. On the contrary, a non-zero imaginary part of $qa$ (named the attenuation constant $\mu=\text{imag}(qa)$) indicates that the elastic wave amplitude will weaken along its propagation from one period to the next.

For the finite sample of the EMM structure, we can also use the transfer matrix method [31] to calculate the vibration frequency response function (FRF) $H(f)$ between the free and excitation edges of the specimen with equation (15).

$$H(f) = \left|\frac{Y(f)}{X(f)}\right| \quad (15)$$

where $Y(f)$ is the vibration on the free edge in frequency domain and $X(f)$ is the one corresponding to the excitation edge.

When the amplitude of the harmonic oscillation of the vibration excitation is set to 1, the calculated FRF equals to $Y(f)$. With equation (15), the vibration FRF can also be measured through vibration experimental instruments. In this paper, vibrations are measured in acceleration instead of displacement. Based on the harmonic assumption, the FRF measured with accelerations are equivalent to that calculated with displacements.

## 2.2. The "pole-zero" designing method

In coordination with the new digital shunting controllers, the expression of $E_p^b(s)$ in equation (6) is regarded as a transfer function where the strain is the input "signal" and the stress is the output one. Thus it can also be described with the typical zero-pole-gain model

$$E_p^b(s) = k \frac{\prod_m (s - z_m)}{\prod_n (s - p_n)} \qquad (16)$$

where $p_n$, $z_m$ and $k$ are system poles, zeros and gain respectively.

Equation (16) can be used to design the equivalent dynamic elastic modulus of PZT patches $b$, and generate LR bandgaps afterwards. A conjugate pair of poles represents a resonant mode. Their imaginary parts are positive/negative angular frequency $\omega_{osc}=2\pi f_{osc}$ of it, while their equal real parts $R$ correspond to the damping factor. When the form of $E_p^b(s)$ is determined, the analog transfer function $H_{DP}(s)$ of the controller can be derived from equation (6). Then the z-transformation are applied to convert $H_{DP}(s)$ into a recurrence equation over discrete times, which is necessary for the programing in microprocessors to fulfil the required $E_p^b(s)$.

**Table 3.** Parameters of transfer function $E_p^b(s)$

| Parameter | Set-1 | Set-2 | Set-3 | Set-4 | Set-5 | Set-6 | Set-7 |
|---|---|---|---|---|---|---|---|
| $R$ | -400 | -350 | -80 | -30 | -300 | -83 | -100 |
|   |   |   |   |   | -400 | -357 | -280 |
|   |   |   |   |   |   |   | -380 |
| $f_{osc}$ (Hz) | 720 | 430 | 210 | 60 | 400 | 185 | 183 |
|   |   |   |   |   | 700 | 420 | 395 |
|   |   |   |   |   |   |   | 680 |
| $\eta$ | 0.5 | 0.5 | 0.5 | 0.5 | 0.3 | 0.5 | 0.2 |
|   |   |   |   |   | 0.8 | 0.6 | 0.7 |
|   |   |   |   |   |   |   | 0.85 |
| $K$ (Pa) | $1\times 10^{11}$ Pa | | | | | | |

All the designed poles, zeros and gain are listed as seven control sets in table 3 and illustrated in figure 2 (the Pole-Zero map). Control sets 1-4 correspond to a single-resonant controller while the others are related to multi-resonant ones (with 2 or 3 resonant modes). As the Young's modulus of the shorted PZT patches used in this paper is about $0.91\times 10^{11}$ Pa[31], the gain is chosen as $k=1\times 10^{11}$ Pa in all the control sets. Each control set of the controller corresponds to one (single-resonant controller) or more (multi-resonant controller) pairs of poles. Each pair of poles corresponds to one pair of zeros with the same real parts $R$ and smaller imaginary parts $\pm\eta\omega_{osc}$, where $\eta$ is regarded as a parameter in the control sets.

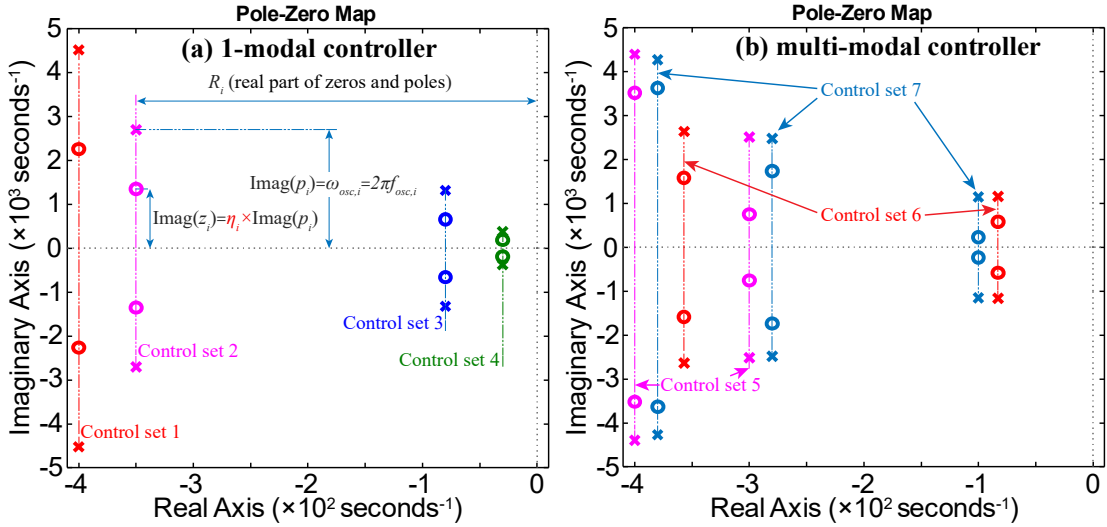

Figure 2. Designed poles and zeros of the transfer function $E_p^b(s)$ for (a) single-resonant and (b) multi-resonant control sets. The crosses and circles represent the poles and zeros respectively.

*2.3. Experimental setups*

Figure 3 shows the vibration experimental set-up as well as the smart EMM specimen. The specimen is made up of an aluminum alloy (6061-T6) beam and 6 arrays of pairs of PZT (P-42) patches glued on it. The lattice constant of the EMM structure $a$=0.1m. All the geometric and material parameters of the specimen can be found in tables 1 and 2. The specimen is hung up by thin and soft strings in order to simulate a free boundary condition. A filtered (1.2 kHz low-pass) white noise signal is generated by the computer and amplified by the power amplifier (HEAS-20). It is then transformed into vibration and applied on left edge of the specimen through the exciter (HEV-20). Vibration accelerations at both the left and right sides of the specimen are measured by two accelerometers (LC0101), a signal conditioner (LC0201-2) and a data acquisition board (USB-1616FS). Post processing of all the phase and amplitude data of the vibration responses is conducted in the computer, thus the vibration FRF from left to the right sides of the specimen is measured. As the circuit board illustrated in figure 1(c) has two independent digital controllers. Three circuit boards are used in the experiments in order to provide the active shunting for six pairs of PZT patches. The circuit boards are powered by a DC power supply (Zhaoxin RXN-6050D) which provides 2 voltage sources (±60 Volts and +5

Volt).

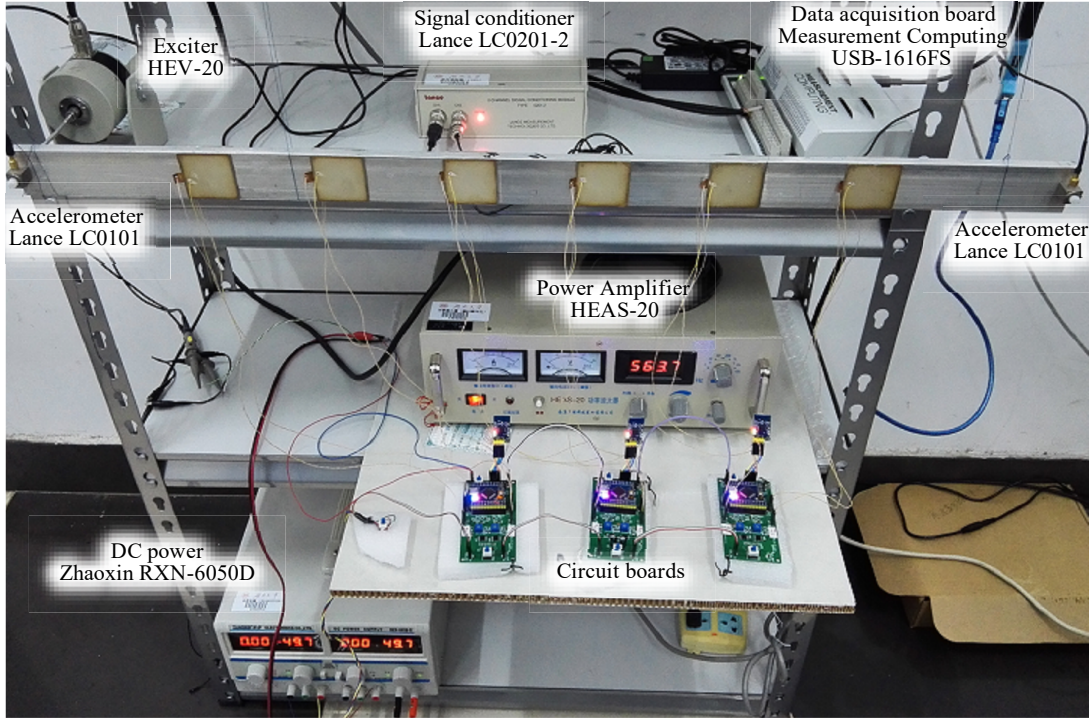

Figure 3. The experimental set-up.

## 3. Results and discussion

In figure 4, the calculated attenuation factors of the smart EMM structure governed by single-resonant controllers (control sets 1-4) are compared with both the calculated and measured vibration FRF of its specimen. The ticks on upper horizontal axis in each sub-figures denotes the normalized frequencies $fa/c_{beam}$, where $f$ is the actual frequency, $a$ is the lattice constant and $c_{beam}$ is the velocity of flexural wave on the hosting beam. The normalized frequency of subwavelength scale should be obviously below 0.5 where the lowest Bragg bandgap exists around. For the hosting beam with rectangle cross section in this paper, $c_{beam}$ can be calculated with[35]

$$c_{beam} = \sqrt{\frac{h\omega}{2\sqrt{3}}}\sqrt{\frac{E_{Al}}{\rho_{Al}}} \tag{17}$$

where $E_{Al}$ is the Young's modulus of the aluminum alloy, $h$ is the thickness of the beam and $\omega$ is the angular frequency.

12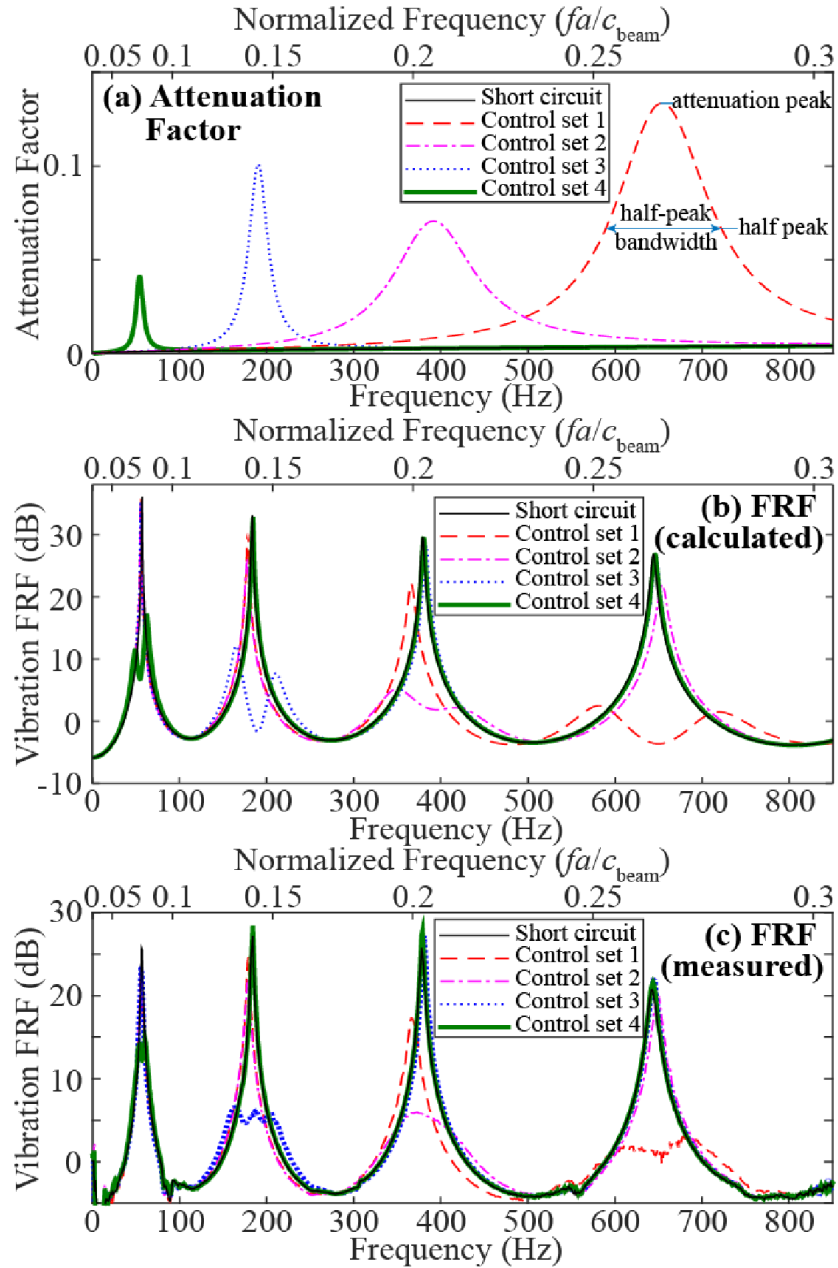

Figure 4. (a) Calculated attenuation factors of the proposed smart EMM governed by single-resonant controller (control sets 1-4). (b) Calculated and (c) measured FRF of the specimen of the smart EMM structure. The thin solid lines represents the results that all the PZT patches are shorted.

All the theoretical and experimental results in figure 4 match basically. Large peaks of attenuation factors corresponding to the designed frequencies are observed in figure 4(a). The normalized frequencies of these attenuation peaks are 0.077, 0.144, 0.206 and 0.267 respectively, which certifies the LR mechanism of these bandgaps. Since attenuation factors reflect the weakening effects of wave propagation per period in an ideal infinite periodic structure, these peaks strongly weaken the corresponding modal peaks in the FRF curves illustrate in figures 4(b) and 4(c). In the experimental results shown in figure 4(c), the lowest 4 modal peaks drop by 10.9dB, 20.9dB, 19.8dB and 18.5dB respectively. In order to evaluate the attenuation abilities more objectively, table 4 illustrates both the calculated and measured average vibration attenuations that induced by the proposed smart EMM for different control sets around different modal frequencies $f_n$. The averaging window in frequency domain is chosen as 90%×$f_n$ to 110%×$f_n$, and those modal frequencies $f_n$ are obtained from the peaks of the FRF curve in the short circuit case. The average attenuation over all the studied frequency range (0-800Hz) are also illustrated in table 4. Those measured attenuations are significant even though they are not as large as the predictions, which is mainly due to the signal–noise limitation of the measurement system where the strongly weakened vibration signal is covered up by noise.

Table 4. Average vibration attenuations of the proposed smart EMM

| | Modal Frequency $f_n$ (Hz) | 57.3 | 183.4 | 379.1 | 644.0 |
|---|---|---|---|---|---|
| **Calculated** average attenuation near modal frequency (dB) $0.9f_n \sim 1.1f_n$ ±10% bandwidth | Control set 1 | 0.41 | 0.73 | 1.75 | **9.17** |
| | Control set 2 | 0.50 | 1.38 | **7.55** | 0.63 |
| | Control set 3 | 0.62 | **11.01** | 0.23 | 0.00 |
| | Control set 4 | **9.29** | 0.04 | -0.01 | -0.03 |
| | Control set 5 | 0.76 | 1.64 | **7.00** | **5.55** |
| | Control set 6 | 0.74 | **5.29** | **7.13** | 0.55 |
| | Control set 7 | 0.90 | **5.71** | **5.20** | **3.87** |
| **Measured** average attenuation near modal frequency (dB) $0.9f_n \sim 1.1f_n$ ±10% bandwidth | Control set 1 | 1.68 | 0.64 | 1.98 | **4.69** |
| | Control set 2 | 0.81 | 1.13 | **4.72** | 0.21 |
| | Control set 3 | 0.69 | **7.40** | -0.05 | -0.13 |
| | Control set 4 | **3.89** | -0.09 | -0.17 | -0.11 |
| | Control set 5 | 1.57 | 1.66 | **4.45** | **2.99** |
| | Control set 6 | 0.40 | **5.46** | **4.34** | 0.15 |
| | Control set 7 | 1.93 | **5.29** | **3.29** | **1.97** |
| Average attenuation within 0-800 Hz (dB) | Control set 1 | 1.42(Calculated) / 0.97(Measured) | | | |
| | Control set 2 | 0.72(Calculated) / 0.44(Measured) | | | |
| | Control set 3 | 0.26(Calculated) / 0.15(Measured) | | | |
| | Control set 4 | 0.01(Calculated) / -0.04(Measured) | | | |
| | Control set 5 | 1.35(Calculated) / 0.89(Measured) | | | |
| | Control set 6 | 0.77(Calculated) / 0.51(Measured) | | | |
| | Control set 7 | 1.08(Calculated) / 0.82(Measured) | | | |

Figure 5(a) illustrates the dispersion curve and attenuation factors of the smart EMM structure when control set-3 is applied, and figure 5(b) demonstrates the equivalent dynamic elastic modulus of beam segments A in the same situation. The subfigures on the right illustrate the corresponding zoom views of figures 5(a) and 5(b). An obvious zigzag on the dispersion curve exists around the attenuation peak with normalized frequency of 0.144. This phenomena indicates negative group velocities around the band gap, which benefits the localization of waves. This region corresponds to a large decrease of the real part the equivalent Young's modulus of beam segments A(with PZT patches and controllers), while different with the frequency where the maximum damping modulus (imaginary part of the Young's modulus) occurs. A narrow Bragg bandgap with smaller attenuation factors can also be found near normalized frequency 0.5.

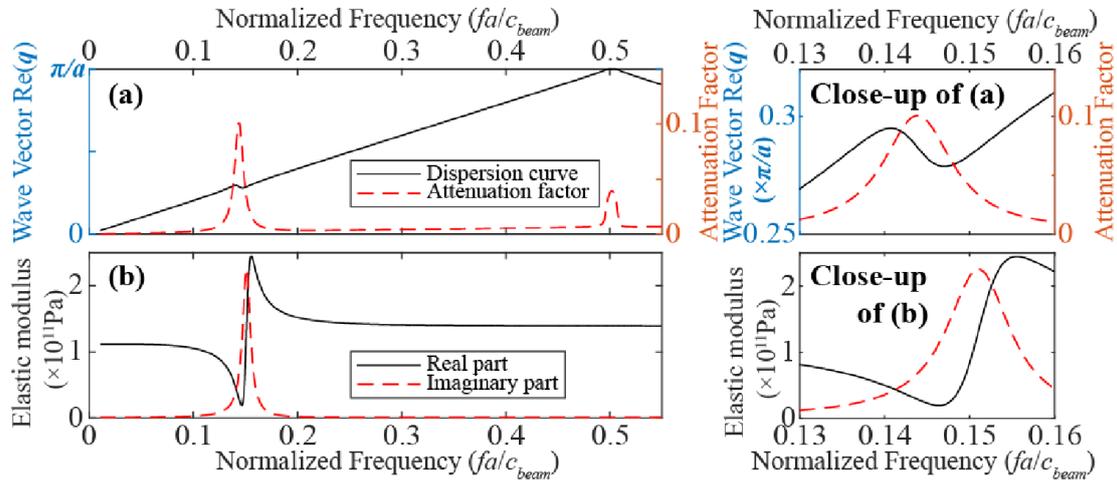

Figure 5. Calculated (a) dispersion curve/attenuation factors and (b) the equivalent elastic modulus of beam segments A when the control set 3 is applied. The subfigures on the right illustrate the corresponding zoom views of figures (a) and (b).

Figure 6 shows how the attenuation peaks and half-peak bandwidth (illustrated in figure 4(a)) change with parameters $f_{osc}$, $R$ and $\eta$. For easy stability check[30], we plot shadow regions where the real part of equivalent Young's modulus of beam segments A is negative. Instead of the frequency of the attenuation peak, the half-peak bandwidth narrows visibly with $R$, and the maximum attenuation factor increases exponentially with $R$. These mean that the parameter $R$ has only the damping effect as previously discussed. Afterwards, when the parameter $\eta$ increases towards 1, the attenuation peak value decreases dramatically while its frequency increases slightly towards the $f_{osc}$. Thus the parameter $\eta$ can be used to adjust the intensity of the attenuation. Finally, when the resonant frequency $f_{osc}$ of the controller increases, both the frequency and value of the attenuation peak increases. Therefore, $f_{osc}$ should be chosen firstly to adjust the frequency of attenuation roughly, and then $\eta$ and $R$ can be used to adjust the intensity and band-width of the attenuation peaks. During the adjusting process, negative value of the equivalent Young's modulus of beam segments A should be avoided in order to insure the stability of the active control system.

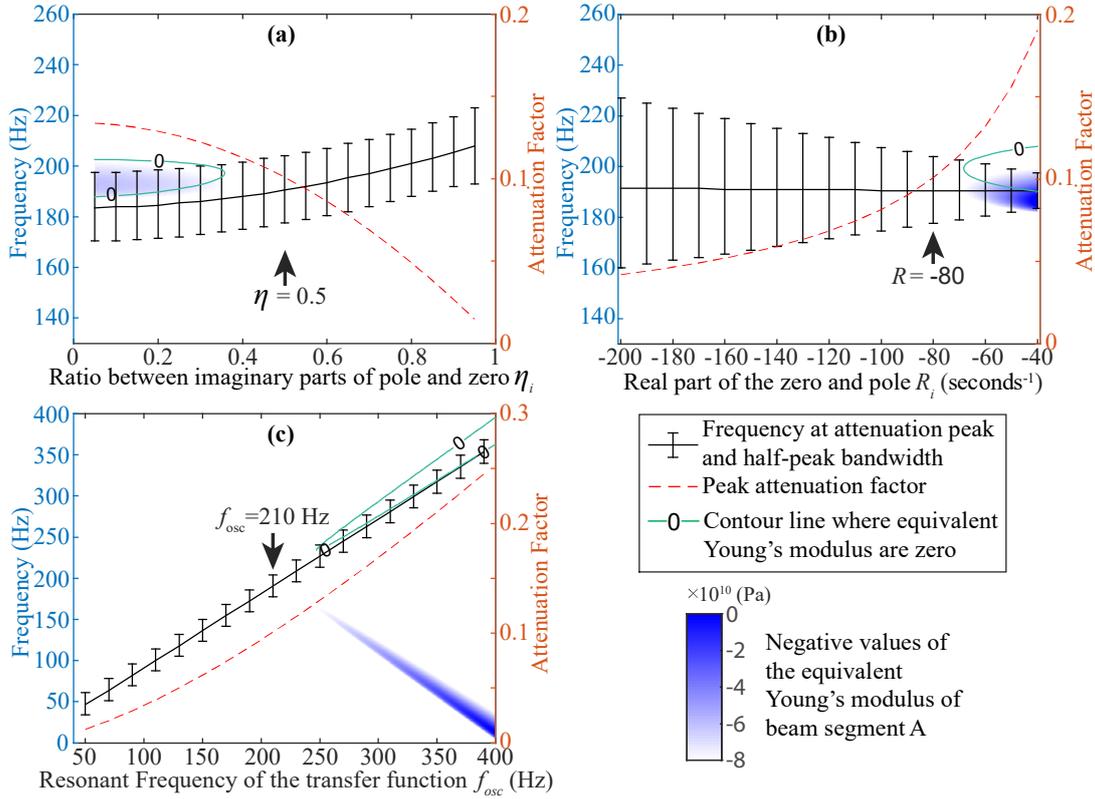

Figure 6. The change of the attenuation peak and half-peak bandwidth with the three parameters (a) $f_{osc}$, (b) $R$ and (c) $\eta$ in the controller. The solid and dashed lines represent the frequency and value of the attenuation peak. Each vertical line with two short ticks on both ends denotes the half-peak bandwidth. The shadow regions with different colors represents where the real part of equivalent Young's modulus of beam segments A is negative. The marking with arrow on each sub-figure shows the parameter which is used in other sub-figures.

When more pairs of poles and zeros are involved in equation (16), multi-resonant controllers can be realized easily instead of the usage of numerous analog electronic component [14]. In figure 7, the calculated attenuation factors of the smart EMM structure governed by dual- and tri-resonant controller (control sets 5-7) are compared with the corresponding calculated and measured vibration FRF. The using of dual- and tri-resonant controllers allows additional low-frequency bandgaps, which is due to the enhanced shunting effect that correspond to the resonances of $E_p^b(s)$. The 3rd-4th, 2nd-3rd or 2nd-4th modal peaks in FRF are strongly and simultaneously attenuated when control sets 5-7 are applied respectively.

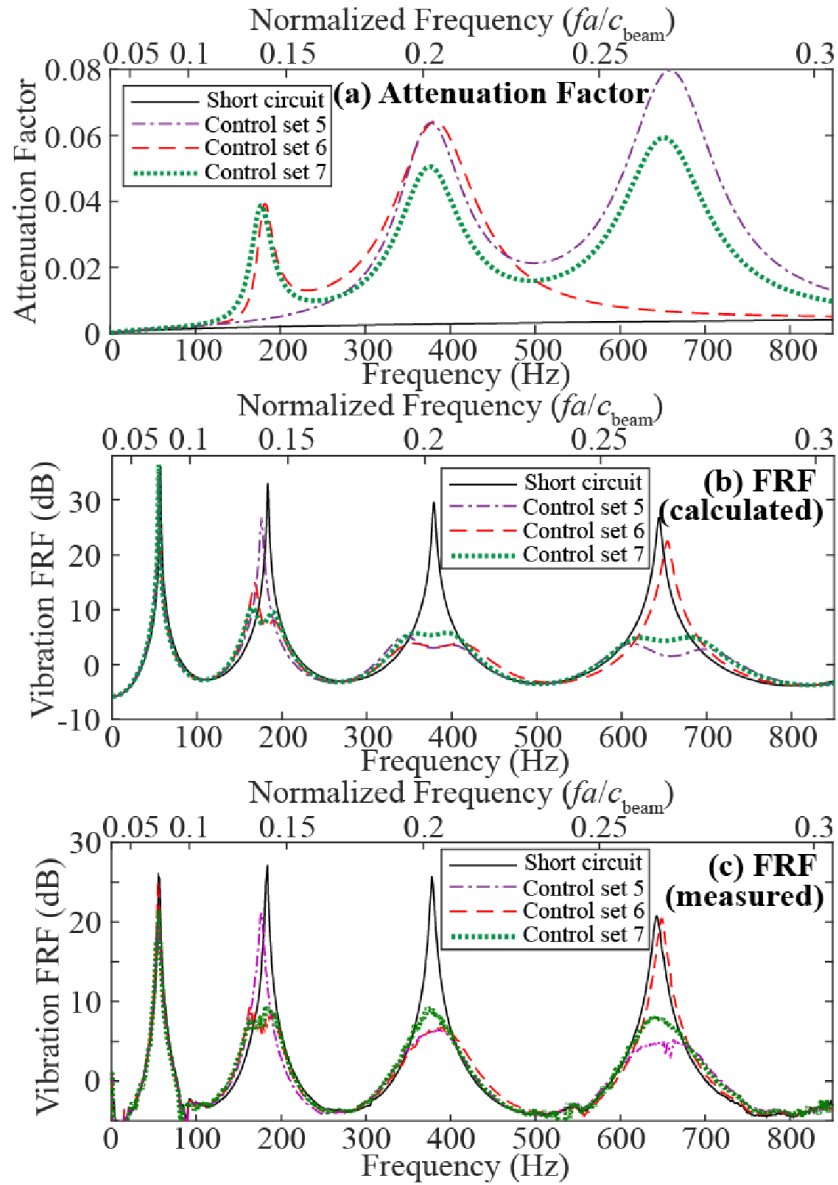

Figure 7. (a) Calculated attenuation factors of the proposed smart EMM governed by multi-resonant controller (control sets 5-7). (b) Calculated and (c) measured FRF of the specimen of the smart EMM structure. The thin solid lines represents the results that all the PZT patches are shorted.

In order to demonstrate how does the number of active-shunted unit-cells affect the ability of attenuations within band gaps, figure 8 illustrates the measured FRF of the smart EMM structure under control set 5 when only 2, 4 or 6 smart shunting units near the vibration exciter are active. All the PZT patches corresponding to the inactive shunting units are shorted. The thin solid lines represents the results that all the PZT patches are shorted. We can see that vibration attenuations with in the band gaps are larger when more cells (or periods) are engaged, which matches with the common sense of metamaterials. When more cells (or periods) are engaged, the properties of the metamaterials are more close to the dispersion analysis based on one unit-cell with Bloch boundary condition. As for the vibration attenuation, more cells ordinarily produce larger attenuations within the bandgaps.

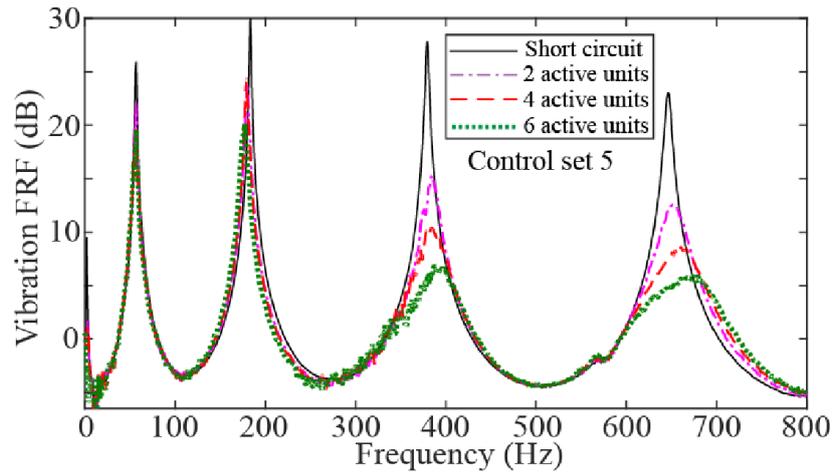

Figure 8. Measured FRF of the smart EMM specimen under control set 5 when only 2, 4 or 6 smart shunting units near the vibration exciter are active. All the PZT patches corresponding to the inactive shunting units are shorted. The thin solid lines represents the results that all the PZT patches are shorted.

4. **Conclusions**

In this paper, we have proposed digital controllers for active piezoelectric shunting and applied them on an elastic metamaterial beam integrated with arrays of piezoelectric patches. The corresponding "pole-zero" design method has been developed, which allows us to design the frequencies, bandwidths and the attenuations of single- or even multi-resonant bandgaps in deep subwavelength region simply by adjusting poles and zeros in the transfer function of Young's modulus. Theoretical and experimental results have shown that the proposed smart metamaterial can achieve a large insulation of up to 20dB in low-frequency ranges. Negative group velocity of flexural wave have been found in the bandgaps, thus the locally resonant mechanism has been confirmed. Furthermore, as parameters in digital controllers can be readily adapted synchronously via wireless broadcasting, the bandgaps of the metamaterial can be tuned easily instead of modifying the structure or the shunting circuits.

**Acknowledgments**

This work was supported by the National Natural Science Foundation of China (Grant Nos. 51322502 and 61501483).

**References**

[1] Liu Z, Zhang X, Mao Y, Zhu Y Y, Yang Z, Chan C T and Sheng P 2000 Locally resonant sonic materials *Science* **289** 1734–6

[2] Wang G, Wen X, Wen J and Liu Y 2006 Quasi-One-Dimensional Periodic Structure with Locally Resonant Band Gap *J. Appl. Mech.-T ASME* **73** 167

[3] Xiao Y, Wen J, Huang L and Wen X 2014 Analysis and experimental realization of locally resonant phononic plates carrying a periodic array of beam-like resonators *J. Phys. D: Appl. Phys.* **47** 45307

[4] Sui N, Yan X, Huang T-Y, Xu J, Yuan F-G and Jing Y 2015 A lightweight yet sound-proof honeycomb acoustic metamaterial *Appl. Phys. Lett.* **106** 171905

[5] Xiao S, Ma G, Li Y, Yang Z and Sheng P 2015 Active control of membrane-type acoustic metamaterial by electric field *Appl. Phys. Lett.* **106** 91904

[6] Cheng Y, Zhou C, Yuan B G, Wu D J, Wei Q and Liu X J 2015 Ultra-sparse metasurface for high reflection of low-frequency sound based on artificial Mie resonances *Nat. Mater.* **14** 1013–9

[7] Zhang H, Xiao Y, Wen J, Yu D and Wen X 2015 Flexural wave band gaps in metamaterial beams with membrane-type resonators: theory and experiment *J. Phys. D: Appl. Phys.* **48** 435305

[8] Zhu R, Liu X N, Hu G K, Sun C T and Huang G L 2014 Negative refraction of elastic waves at the deep-subwavelength scale in a single-phase metamaterial *Nat. Commun.* **5** 5510

[9] Zhou X and Hu G 2011 Superlensing effect of an anisotropic metamaterial slab with near-zero dynamic mass *Appl. Phys. Lett.* **98** 263510

[10] Zhu J, Christensen J, Jung J, Martin-Moreno L, Yin X, Fok L, Zhang X and Garcia-Vidal F J 2011 A holey-structured metamaterial for acoustic deep-subwavelength imaging *Nat. Phys.* **7** 52–5

[11] Cummer S A, Popa B-I, Schurig D, Smith D R, Pendry J, Rahm M and Starr A 2008 Scattering Theory Derivation of a 3D Acoustic Cloaking Shell *Phys. Rev. Lett.* **100** 24301

[12] Thorp O, Ruzzene M and Baz A 2001 Attenuation and localization of wave propagation in rods with periodic shunted piezoelectric patches *Smart Mater. Struct.* **10** 979–89

[13] Airoldi L and Ruzzene M 2011 Design of tunable acoustic metamaterials through periodic arrays of resonant shunted piezos *New J. Phys.* **13** 113010

[14] Airoldi L and Ruzzene M 2011 Wave Propagation Control in Beams Through Periodic Multi-

Branch Shunts *J. Intell. Mater. Syst. Struct.* **22** 1567–79

[15]   Wang G, Chen S and Wen J 2011 Low-frequency locally resonant band gaps induced by arrays of resonant shunts with Antoniou's circuit: experimental investigation on beams *Smart Mater. Struct.* **20** 15026

[16]   Chen S and Wang G 2016 Wave propagation in beams with anti-symmetric piezoelectric shunting arrays *Chin. Phys. B* **25** 34301

[17]   Spadoni A, Ruzzene M and Cunefare K 2009 Vibration and wave propagation control of plates with periodic arrays of shunted piezoelectric patches *J. Intell. Mater. Syst. Struct.* **20** 979–90

[18]   Casadei F, Ruzzene M, Dozio L and Cunefare K A 2010 Broadband vibration control through periodic arrays of resonant shunts: experimental investigation on plates *Smart Mater. Struct.* **19** 15002

[19]   Casadei F, Beck B S, Cunefare K A and Ruzzene M 2012 Vibration control of plates through hybrid configurations of periodic piezoelectric shunts *J. Intell. Mater. Syst. Struct.* **23** 1169–77

[20]   Chen S, Wang G, Wen J and Wen X 2013 Wave propagation and attenuation in plates with periodic arrays of shunted piezo-patches *J. Sound Vib.* **332** 1520–32

[21]   Wen J, Chen S, Wang G, Yu D and Wen X 2016 Directionality of wave propagation and attenuation in plates with resonant shunting arrays *J. Intell. Mater. Syst. Struct.* **27** 28–38

[22]   Zhang H, Wen J, Xiao Y, Wang G and Wen X 2015 Sound transmission loss of metamaterial thin plates with periodic subwavelength arrays of shunted piezoelectric patches *J. Sound Vib.* **343** 104–20

[23]   Casadei F, Delpero T, Bergamini A, Ermanni P and Ruzzene M 2012 Piezoelectric resonator arrays for tunable acoustic waveguides and metamaterials *Journal of Applied Physics* **112** 64902

[24]   Kwon B-J, Jung J-Y, Lee D, Park K-C and Oh I-K 2015 Tunable acoustic waveguide based on vibro-acoustic metamaterials with shunted piezoelectric unit cells *Smart Materials and Structures* **24** 105018

[25]   Nouh M A, Aldraihem O J and Baz A 2015 Periodic metamaterial plates with smart tunable local resonators *Journal of Intelligent Material Systems and Structures*

[26]   Tateo F, Collet M, Ouisse M, Ichchou M, Cunefare K and Abbe P 2014 Experimental characterization of a bi-dimensional array of negative capacitance piezo-patches for vibroacoustic control *J. Intell. Mater. Syst. Struct.* **26** 952–64

[27]   Zhu R, Chen Y Y, Barnhart M V, Hu G K, Sun C T and Huang G L 2016 Experimental study of an adaptive elastic metamaterial controlled by electric circuits *Appl. Phys. Lett.* **108** 11905

[28]   Celli P and Gonella S 2015 Tunable directivity in metamaterials with reconfigurable cell symmetry *Applied Physics Letters* **106** 91905

[29]   de Marneffe B and Preumont A 2008 Vibration damping with negative capacitance shunts: theory and experiment *Smart Mater. Struct.* **17** 35015

[30]   Wang G, Wang J, Chen S and Wen J 2011 Vibration attenuations induced by periodic arrays of piezoelectric patches connected by enhanced resonant shunting circuits *Smart Mater. Struct.* **20** 125019

[31]   Wang G and Chen S 2016 Large low-frequency vibration attenuation induced by arrays of piezoelectric patches shunted with amplifier–resonator feedback circuits *Smart Mater. Struct.* **25** 15004

[32]   Zhou W, Wu Y and Zuo L 2015 Vibration and wave propagation attenuation for metamaterials by periodic piezoelectric arrays with high-order resonant circuit shunts *Smart Mater. Struct.* **24** 65021

[33]   Méndez-Sánchez R, Morales A and Flores J 2005 Experimental check on the accuracy of Timoshenko's beam theory *Journal of sound and vibration* **279** 508–12

[34]   Kittel C 1986 *Introduction to solid state physics* (New York: Wiley)

[35]   Wang G, Wen J and Wen X 2005 Quasi-one-dimensional phononic crystals studied using the improved lumped-mass method: Application to locally resonant beams with flexural wave band gap *Phys. Rev. B* **71** 104302